\documentclass[11pt]{article}
\usepackage{moriond,epsfig}

\makeatletter
\def\@cite#1#2{{[{#1}]\if@tempswa\typeout
{IJCGA warning: optional citation argument
ignored: `#2'} \fi}}

\newcount\@tempcntc
\def\@citex[#1]#2{\if@filesw\immediate\write\@auxout{\string\citation{#2}}\fi
  \@tempcnta\z@\@tempcntb\m@ne\def\@citea{}\@cite{\@for\@citeb:=#2\do
    {\@ifundefined
       {b@\@citeb}{\@citeo\@tempcntb\m@ne\@citea\def\@citea{,}{\bf ?}\@warning
       {Citation `\@citeb' on page \thepage \space undefined}}%
    {\setbox\z@\hbox{\global\@tempcntc0\csname b@\@citeb\endcsname\relax}%
     \ifnum\@tempcntc=\z@ \@citeo\@tempcntb\m@ne
       \@citea\def\@citea{,}\hbox{\csname b@\@citeb\endcsname}%
     \else
      \advance\@tempcntb\@ne
      \ifnum\@tempcntb=\@tempcntc
      \else\advance\@tempcntb\m@ne\@citeo
      \@tempcnta\@tempcntc\@tempcntb\@tempcntc\fi\fi}}\@citeo}{#1}}
\def\@citeo{\ifnum\@tempcnta>\@tempcntb\else\@citea\def\@citea{,}%
  \ifnum\@tempcnta=\@tempcntb\the\@tempcnta\else
   {\advance\@tempcnta\@ne\ifnum\@tempcnta=\@tempcntb \else \def\@citea{--}\fi
    \advance\@tempcnta\m@ne\the\@tempcnta\@citea\the\@tempcntb}\fi\fi}
\makeatother

\def\be{\begin{equation}}
\def\ee{\end{equation}}
\def\bea{\begin{eqnarray}}
\def\eea{\end{eqnarray}}

\def\lsim{\mathrel{\raise.3ex\hbox{$<$\kern-.75em\lower1ex\hbox{$\sim$}}}}
\def\gsim{\mathrel{\raise.3ex\hbox{$>$\kern-.75em\lower1ex\hbox{$\sim$}}}}
\def\amu{\delta a_\mu^{\rm NP}}
\def\sinbma{\sin(\beta-\alpha)}
\def\tanb{\tan\beta}
\def\hl{h}
\def\ha{A}
\def\hh{H}
\def\hpm{H^\pm}
\def\mha{m_{\ha}}
\def\mhl{m_{\hl}}
\def\mhh{m_{\hh}}
\def\mhpm{m_{\hpm}}

\def\ifmath#1{\relax\ifmmode #1\else $#1$\fi}
\def\ls#1{\ifmath{_{\lower1.5pt\hbox{$\scriptstyle #1$}}}}
\begin{document}
\vspace*{4cm}
\title{A Light Higgs Boson Explanation for the $g-2$ Crisis \footnote{Talk
given at the XXXVIth Rencontres de Moriond  
``Electroweak Interactions and Unified Theories'',
Les Arcs 1800,  March 10-17, 2001.  }}

\author{ A. Dedes$^1$ and H. E. Haber$^2$ }

\address{$^1$Physikalisches Institut der Universit\"at Bonn,\\ 
 Nu\ss allee 12, D-53115 Bonn, Germany}

\address{$^2$Santa Cruz Institute for Particle Physics\\ 
University of California, Santa Cruz, CA, 95064 USA }

\maketitle\abstracts{A light CP-even Higgs
boson with a mass of around 10 GeV could explain the recent 
BNL measurement of the muon anomalous magnetic moment. This observation
is based on
a general CP-conserving two Higgs 
doublet extension of the Standard Model with 
no tree-level 
flavor changing neutral current couplings.  The
Higgs mass is constrained by experiments at CESR and LEP 
to be less than twice the lightest B-meson mass
and greater than (roughly) the Upsilon mass.  It may be
possible to exclude or discover such a Higgs boson by fully analyzing 
the existing LEP data.}

\section{A possibility for New Physics from the $g-2$ measurement at BNL E821}
Based on Davier-H\"ocker calculation~\cite{hocker} 
for the hadronic contributions
to the muon anomalous magnetic moment, BNL~\cite{bnl}
has reported a 2.6$\sigma$ deviation
from the Standard Model (SM) prediction. 
There are four possible interpretations
of this discrepancy~\cite{ond,marciano} :
\begin{itemize}

\item a statistical fluctuation with 0.9\% probability.

\item something is ``wrong'' with the experiment : this
is going to  be resolved within the coming year~\cite{ond} since 
the data from the 2000 and 2001 runs have yet to be analyzed.

\item something is ``wrong'' with the theoretical analysis of the
hadronic contribution to
the muon anomalous magnetic moment, $a_\mu^{\rm had}$. 
This analysis depends critically on the experimental data
for $\sigma(e^+e^-\to$~hadrons) below $\sqrt{s}=2$~GeV.
One expects forthcoming data from Novorsibirsk and BES, and 
an intense theoretical effort from a number of 
groups~\cite{hocker,erler,ynd,narison,jerg}
to reduce the systematic uncertainties that contribute to $a_\mu^{\rm had}$.
[It is 
important to note here that the error on $a_\mu^{\rm had}$ is
necessarily larger when not using tau data. Indeed, 
two~\cite{hocker,narison} of the
recent numbers have significantly been improved by the use of hadronic
tau data  (in the two and four pion final states) in conjunction with
the conserved vector current. New results on $\tau^+ \rightarrow \pi^+ \pi^0 
\nu $ decays are expected from ALEPH soon~\cite{privandrea}.]

\item the by far most exciting possibility : there is a 
contribution due to New Physics
beyond the SM, $\amu$. Based on two different analyses of the
hadronic contributions, one finds at 90\% CL~\cite{hocker,narison} 
\begin{eqnarray}
170 \times 10^{-11} \lsim &\amu & \lsim 690 \times 10^{-11}~~~ 
[{\rm Davier-Hocker (98)}]\;, \label{dh}\\
97 \times 10^{-11} \lsim &\amu & \lsim 667 \times 10^{-11}~~~ 
[{\rm Narison (01)}]\,. \label{nar}
\end{eqnarray}

\end{itemize}
In what follows, we assume  the fourth possibility as an explanation
to the discrepancy under discussion. Two remarks concerning
eqs.~(\ref{dh}) and (\ref{nar})
are in order~\cite{marciano} :
{\it (i)} the contribution from the New Physics is positive, $\amu >0$,
and {\it (ii)},  is of the order of the Electroweak (EW) contributions,
{\it i.e.,} $\amu \propto G_F m_\mu^2/(4\pi^2\sqrt{2})$.     

In the SM, the 1-loop Higgs boson contribution to the muon anomalous 
magnetic moment is suppressed by a factor of $m_\mu^2/m_h^2$.  This
factor is particularly small given that the
SM Higgs mass bound from LEP is $m_h> 113.5$~GeV.
The situation is potentially different if we extend the SM by adding an extra 
Higgs doublet~\cite{hunters}. The idea that a light Higgs boson 
could enhanced the predicted value of $a_\mu$ has been recently 
employed in~\cite{dedhab,kraw}. Here we strictly follow the 
discussion of~\cite{dedhab}~\footnote{For the purposes of this presentation
we refer to the Two Higgs Doublet Model (II)
[2HDM(II)]~\cite{hunters}.
 For the Model~I
see the discussion in~\cite{dedhab} and for Model~III see~\cite{2hdm3}.}.
 The enhancement arises from the Higgs sector is twofold:
{\it (i)} the coupling of the muon  to one of the CP-even Higgs 
bosons ($h$ or $H$),
 CP-Odd Higgs ($A$) and charged higgs ($H^\pm$)
 is proportional to $\tan\beta$ (the ratio
of the two vacuum expectation values), and hence $\amu \propto
\tan^2\beta$;
 {\it (ii)} the Higgs coupling
to the $Z$-boson  can be set to zero, {\it i.e.,} $\sinbma=0$, 
(where $\alpha$
is the CP-even Higgs bosons mixing angle)
and thus LEP constraints from
$e^+e^-\rightarrow hZ$
do not apply~\cite{sin2}. In this case, as we show below, the Higgs
mass can be as light as $\sim 10$ GeV.

\section{Model~II Higgs boson corrections to $a_\mu$}

The first calculation of the one-loop electroweak corrections to the muon
anomalous magnetic moment was presented by Weinberg and
Jackiw~\cite{Jackiw:1972jz} and 
by Fujikawa, Lee and Sanda~\cite{Fujikawa:1972fe}.  A very useful
compendium of formulae for the one-loop corrections to $g-2$
in a general electroweak model was
given in~\cite{Leveille:1978rc}, 
and applied to the 2HDM in~\cite{haber}.
In the 2HDM, both neutral and charged Higgs bosons contribute to $g-2$.
 One can derive approximate results 
by expanding the loop integrals in terms of the parameters
$m_\mu^2/m_{h,H,A,H^\pm}^2$.


At the end of 
section 1, we argued that the most significant Higgs contribution
to $\amu$ (consistent with the LEP SM Higgs search) arises in the
parameter regime in which $\sinbma\simeq 0$ and $\tanb\gg 1$.  
Setting  $\sinbma=0$ and keeping only the leading terms in 
$m_\mu^2/m_{h,H,A,H^\pm}^2$ when evaluating the above integrals, 
the total Higgs sector
 contribution to $a_\mu$ at 1-loop\footnote{All 
the results given here are based on a one loop calculation.  The
authors of~\cite{2loop} argue that significant two-loop
contributions contribute to $\delta a_\mu^{\rm Higgs}$ due primarily
to the effect of one particular [Barr-Zee] diagram, which is of the
same order as the corresponding one-loop contribution, but with
opposite sign.  Adding this contribution to that of eq.~(\ref{hfull})
therefore reduces the result for $\delta a_\mu^{\rm Higgs}$ at fixed
$m_h$, implying that $m_h\lsim m_b$ in order that the total Higgs
contribution to $a_\mu$ lie within the range suggested by
eqs.~(\ref{dh}) and (\ref{nar}).  However, this possibility is
excluded by the non-observation of $\Upsilon\to h\gamma$ at CESR.
By using the results of~\cite{chang} for the two loop contributions
from the CP-Odd Higgs boson to the g-2, reference~\cite{cpodd} 
argues that the two-loop contribution of the
Barr-Zee diagram could change the sign of the one-loop Higgs contribution, in
which case a light CP-odd Higgs boson could generate an overall {\it
positive} contribution to $\delta a_\mu$ of the required magnitude.
However, there is at present no complete two-loop computation of the
electroweak contribution to $\delta a_\mu$ in the 2HDM.  Whether the 
dominance of the Barr-Zee diagram survives a more complete two-loop
analysis remains to be seen.}
is given by:
\begin{eqnarray}
&& \delta a_\mu^{\rm Higgs}=  \delta a_\mu^h+\delta a_\mu^H
+\delta a_\mu^A+
\delta a_\mu^{H^\pm}  \nonumber \\[5pt]
&& \qquad\simeq\frac{G_Fm_\mu^2}{4\pi^2 \sqrt{2}}
\tan^2\beta \bigg \{
\frac{m_\mu^2}{\mhl^2} \biggl[\ln\left(\frac{\mhl^2}{m_\mu^2}\right)
-\frac{7}{6} \biggr] -\frac{m_\mu^2}{\mha^2} \biggl[
\ln\left(\frac{\mha^2}{m_\mu^2}\right) -\frac{11}{6}\biggr]-
\frac{m_\mu^2}{6m_{H^{\pm}}^2} \biggr \} \;. 
\label{hfull}
\end{eqnarray}
Note that the 
logarithms appearing in eq.(\ref{hfull}) always dominate the corresponding
constant terms when the Higgs masses are larger than 1 GeV.
It is then clear that $\ha$ and $\hpm$ exchange contribute a
negative value to $\amu$ and thus cannot explain   the BNL $g-2$
measurement which suggests a positive value for $\amu$. In addition,
we  should take
$\mha$ and $\mhpm$ large (masses above 100~GeV are sufficient)
in order that the corresponding
$\ha$ and $\hpm$ negative contributions are 
neglibly small. If $\amu$ is to 
be a consequence of the Higgs sector, it must be entirely
due to the contribution of the light CP-even Higgs boson.
Note that the heavier
CP-even Higgs, $\hh$, does not give a contribution proportional to
$\tan\beta$ ; hence its contribution
to $\amu$ can be neglected in eq.(\ref{hfull}). 
Thus, to a good approximation,

\begin{equation}
\delta a_{\mu}^{\rm Higgs}\simeq 
\delta a_\mu^{{\hl}} \simeq \frac{G_Fm_\mu^2}{4\pi^2 \sqrt{2}}
\left(\frac{m_\mu^2}{\mhl^2}\right) \tan^2\beta
\bigg[
\ln\left(\frac{\mhl^2}{m_\mu^2}\right)-\frac{7}{6} \biggr ] \;.
\label{app1}
\end{equation}

One can check that a light Higgs boson with a mass of
around 10 GeV and with $\tan\beta=35$ gives $\delta a_{\mu}^{\rm
Higgs}\simeq 280 \times 10^{-11}$, which is within the
90\%~CL allowed range for $\amu$ quoted in eqs.~(\ref{dh}) and (\ref{nar}).
Contour lines corresponding to a full numerical 
evaluation of the Higgs sector one-loop 
contribution to $\delta
a_\mu^{\rm Higgs}$ [in units of $10^{-11}$] are exhibited
in fig.~1, for $\sinbma=0$ and $\mhh=\mha=\mhpm=200$~GeV.\footnote{%
The results are insensitive to the values of the heavy Higgs masses
above 100~GeV.}
The relevant experimental bounds are also displayed
in fig.~1; [see discussion below].
A careful inspection of the excluded region in the $\mhl$ {\it vs.}
$\tanb$ parameter space shows that a light
Higgs boson of around 10 GeV mass and $20\lsim \tan\beta \lsim 35$
is permitted depending on the lower bound of eqs.~(\ref{dh}) and (\ref{nar}).
In this parameter regime, we obtain
a value for $\amu$ within the $90\%$~CL allowed range of
eq.~(\ref{dh}).  However, the central values of $\amu$ given
in~\cite{marciano,narison} lie within the excluded regions of fig.~1.

\begin{figure}
\rule{5cm}{0.2mm}\hfill\rule{5cm}{0.2mm}
\vskip 2.5cm
\centerline{\psfig{figure=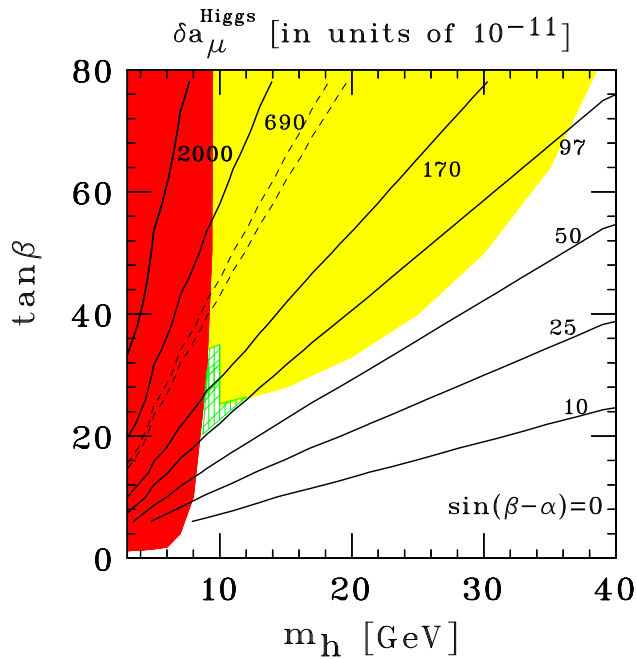,height=3.3in,angle=90}}
\caption{%
Contours of the  predicted one-loop Higgs sector contribution
to the muon anomalous magnetic moment, 
$\delta \alpha_\mu^{\rm Higgs}$ (in units of $10^{-11}$)
in the 2HDM, assuming that $\sin(\beta-\alpha)=0$, 
and $\mhh=\mha=\mhpm=200$~GeV
(there is little sensitivity to the heavier Higgs masses).
The dashed line contours correspond to the central value of
$\delta a_\mu\equiv a_\mu^{\rm exp}-a_\mu^{\rm SM}$, as reported 
by Brown {\it et.al} and by Narison.
The contour lines marked 170 and 690 correspond to 90\%~CL limits
for the contribution of new physics to $a_\mu$ [eq.~(\ref{dh})]. The contour
97 corresponds to the lower bound on $\amu$ given in eq.(\ref{nar}).
The dark-shaded (red) region is excluded 
by the CUSB Collaboration search for $\Upsilon\to\hl\gamma$ 
at CESR.  The light-shaded (yellow) region 
is excluded at $95\%$~CL by the ALEPH and DELPHI searches
for $e^+e^-\to \hl f\bar f$ ($f=b$ or $\tau$) at LEP.
In the small hatched region (green) nestled between the 
two experimentally excluded
shaded regions, above the 97 contour line and 
centered around $\mhl\simeq 10$~GeV, the Higgs sector
contribution to $\amu$ lies within the 90\% CL allowed range.
}
\end{figure}

\section{Experimental constraints from CESR and LEP}

Let us consider the 2HDM in which $\sinbma=0$, $\tanb\gg 1$
and $\mhl\sim {\cal O}(10$~GeV), which are necessary conditions
if the Higgs sector is to be the source for $\amu$ in the range
given by eq.(\ref{dh},\ref{nar}).
The $\hl\ha Z$ coupling is maximal , so we 
must assume that $\mha$ is large enough so that
$e^+e^-\to \hl\ha$ is not observed at LEP.
The tree-level $\hl ZZ$ coupling is absent, which implies that
the LEP SM Higgs search based on $e^+e^-\to Z\to Z\hl$ does not impose
any significant constraints on $\mhl$.\footnote{Presumably,
radiative corrections would lead to a small effective value for
$\sinbma$.  The LEP Higgs search yields an excluded region in the 
$\sinbma$ {\it vs.} $\mhl$ plane, and implies that for $\mhl\sim 10$~GeV, 
$|\sinbma|\lsim 0.06$ is not excluded
at 95\%~CL~\cite{sin2}.}  
However, there are a number of constraints on light Higgs masses 
that do not rely on the $\hl ZZ$ coupling.  For Higgs bosons with
$\mhl\lsim 5$~GeV, the SM Higgs boson was ruled out by a
variety of arguments that were summarized in \cite{hunters}.  For
5~GeV $\lsim\mhl\lsim 10$~GeV, the relevant
Higgs boson constraint can be derived 
from the absence of Higgs production in $\Upsilon\to\hl\gamma$ [see
dark-shaded (red) region in Fig.1].
A second bound on $\mhl$ can be derived from the non-observation of 
Higgs bosons at LEP  via the 
process $e^+e^-\to \hl f\bar f$ ($f=b$, $\tau$). However, only preliminary results have appeared so far
by DELPHI~\cite{delphi} and ALEPH~\cite{aleph} [see 
light-shaded (yellow) region in
Fig.1].
One noteworthy consequence of $\mhl\sim 10$~GeV is
the possibility of mixing
between the $\hl$ and the $0^{++}$ $b\bar b$ bound states
$\chi\ls{b0}(1P)$ and $\chi\ls{b0}(2P)$, as
discussed in \cite{haber,egns}.  As a result, the decay
$\chi\ls{b0}\to\tau^+\tau^-$ should be prominent. For more details
see~\cite{dedhab}.

\section{Results}

Using the experimental bounds on the Higgs mass discussed in
section 3, 
we conclude that a light Higgs boson
can be responsible for the observed
$2.6\sigma$ deviation of the BNL measurement of the muon 
anomalous magnetic moment at the 90\%~CL in the framework of a
CP-conserving~\cite{Wu}
two-Higgs-doublet model with Model~II Higgs-fermion Yukawa couplings 
only if the model parameters satisfy the
following requirements:
\begin{eqnarray}
&& m_{\Upsilon}\lsim \mhl \lsim 2 m_B \;, \nonumber \\
&& \sinbma\simeq 0 \;, \nonumber \\
&& 27\lsim\tanb\lsim 35~~~{\rm [Davier-Hocker (98)]} \;, \nonumber \\
&& 20\lsim\tanb\lsim 35~~~{\rm [Narison (01)]} \;.
\label{result}
\end{eqnarray}
In addition, the other Higgs bosons ($H$, $A$ and $H^\pm$)
should be heavy enough 
to be consistent with the non-observation of $HZ$, $h A$ and
$H^+H^+$ production at LEP.  Although, the parameter space 
is highly constrained, it is still useful to extend the LEP search
for $f\bar f h$ production ($f=b$ or $\tau$) to lower values of
$\tan\beta$ and Higgs mass to either confirm or rule out the 
parameter range specified in eq.~(\ref{result}).

An extended Higgs sector provides one possible source of
new physics that can be probed by a precision measurement of
the muon anomalous magnetic moment.  Many other new physics
mechanisms for explaining the recent BNL measurement have
also been explored~\cite{susy,tecni,lepto,exotic,modelind}.
If the $g-2$ ``crisis'' persists, it will be essential to find 
direct effects of the new physics at future colliders in order to 
establish the origin of a non-SM component to the muon anomalous 
magnetic moment.


\section*{Acknowledgments}

AD would like to thank A. H\"ocker, H. Dreiner, G. Onderwater, M. Schumacher, 
P. Teixeira Dias, W. Murray, R. Teuscher, P. Janot, J.M. Frere, 
E. Tournefier, M. K\"obel and A. Sanda for 
illuminating discussions on the muon $g-2$, Higgs and EW searches, and also
the organizers of this excellent workshop. AD acknowledges also 
the RTN European Program HPRN-CT-2000-00148 for financial support.

\end{document}
